\documentclass[conference]{IEEEtran}
\IEEEoverridecommandlockouts
\usepackage{cite}
\usepackage{amsmath,amssymb,amsfonts}

\usepackage{graphicx}
\usepackage{textcomp}
\usepackage{xcolor}
\usepackage{comment}
\usepackage[ruled,vlined]{algorithm2e}%
\usepackage[flushleft]{threeparttable}%
\usepackage{tabularx}
\usepackage{makecell}

\def\BibTeX{{\rm B\kern-.05em{\sc i\kern-.025em b}\kern-.08em
    T\kern-.1667em\lower.7ex\hbox{E}\kern-.125emX}}
\begin{document}

\title{QMoE: A Quantum Mixture of Experts Framework for Scalable Quantum Neural Networks}

\author{
    Hoang-Quan Nguyen$^1$, Xuan-Bac Nguyen$^1$, Sankalp Pandey$^1$ \\
    Samee U. Khan$^2$, Ilya Safro$^3$, Khoa Luu$^1$ \\
    $^1$Department of Electrical Engineering and Computer Science, University of Arkansas, AR \\
    $^2$Department of Electrical and Computer Engineering, Kansas State University, KS \\
    $^3$Department of Computer and Information Sciences, University of Delaware, DE \\
    \small{\texttt{\{hn016, xnguyen, sankalpp, khoaluu\}@uark.edu}} \quad
    \small{\texttt{sameekhan@ksu.edu}} \quad \small{\texttt{isafro@udel.edu}}
}

\maketitle

\begin{abstract}
Quantum machine learning (QML) has emerged as a promising direction in the noisy intermediate-scale quantum (NISQ) era, offering computational and memory advantages by harnessing superposition and entanglement. However, QML models often face challenges in scalability and expressiveness due to hardware constraints. In this paper, we propose quantum mixture of experts (QMoE), a novel quantum architecture that integrates the mixture of experts (MoE) paradigm into the QML setting. QMoE comprises multiple parameterized quantum circuits serving as expert models, along with a learnable quantum routing mechanism that selects and aggregates specialized quantum experts per input. The empirical results from the proposed QMoE on quantum classification tasks demonstrate that it consistently outperforms standard quantum neural networks, highlighting its effectiveness in learning complex data patterns. Our work paves the way for scalable and interpretable quantum learning frameworks.
\end{abstract}


\section{Introduction}

Quantum computing offers the potential for exponential acceleration in solving specific problems compared to classical computing due to the unique properties of superposition and entanglement \cite{preskill2012quantum,preskill2018Quantum,boixo2018characterizing}.
With the growing number of qubits in the noisy intermediate-scale quantum (NISQ) era, the prospect of practical quantum advantage is becoming more feasible. Among various applications, quantum machine learning (QML) stands out as a promising area, given its high computational demands and tolerance to quantum noise.
Prior studies have introduced QML frameworks analogous to classical approaches, including quantum K-nearest neighbor \cite{basheer2020quantum}, support vector machines \cite{rebentrost2014quantum}, quantum clustering \cite{horn2001method,horn2001algorithm,nguyen2023quantum,nguyen2024qclusformer}, and quantum neural networks (QNNs) \cite{ezhov2000quantum,zhou2023quantum,gupta2020quantum,dendukuri2019defining,dendukuri2018image,nguyen2024hierarchical,nguyen2025diffusion,nguyen2024quantum}.
Leveraging the quantum properties, QML methods offer advantages in both computation and memory efficiency \cite{biamonte2017quantum,du2020expressive}, enabling the handling of large-scale machine learning tasks that may be challenging for classical systems.

The mixture of experts (MoE) paradigm has proven to be an effective strategy for enhancing scalability and specialization in deep learning models. MoE architectures typically consist of multiple expert networks, each trained to specialize on a subset of the input space, and a routing mechanism that dynamically selects a sparse subset of experts per input \cite{shazeer2017outrageously,fedus2022switch}.
This modular structure enables MoE models to achieve high capacity while maintaining computational efficiency, and has recently led to significant performance improvements in large-scale deep neural network systems.

The integration of MoE principles into QML presents an opportunity to address two central challenges in QML problems, including model capacity and resource allocation. While QML models have shown promise, the constraints of the available qubits and circuit depth often limit their expressive power. By adopting an MoE-inspired approach, one can distribute learning across multiple quantum experts, each implemented as a specialized quantum circuit. Meanwhile, a trainable quantum dynamically selects which experts to activate. 
It introduces sparsity, modularity, and specialization into QML, which are crucial for scaling quantum models under hardware limitations.

In this work, we introduce a novel quantum architecture, named quantum mixture of experts (QMoE), that combines the efficiency of classical MoE systems with the expressive power of quantum models. QMoE consists of a set of parallel parameterized quantum circuits serving as experts, and a learned routing function that computes a weighted aggregation of these circuits for each input. Our experiments demonstrate that QMoE can outperform the standard QNNs in classification tasks, highlighting the advantages of QMoE over prior QML models.

\textbf{Contributions of this Work} 
The contributions of this work are threefold. First, we propose quantum mixture of experts, a quantum computing framework for learning and modeling different data patterns. Second, we introduce a quantum routing circuit and parameterized quantum experts to determine the input data for the specialized experts. Lastly, the empirical experimental results demonstrate the novelties and effectiveness of our proposed framework compared to conventional methods.
Through this contribution, we aim to open new directions for scalable and interpretable quantum learning systems.

\section{Background}\label{background}

\subsection{Mixture of Experts (MoE)}

Mixture of experts is a long-established ensemble technique that decomposes complex predictive tasks across multiple specialized sub-networks, referred to as experts \cite{jacobs1991adaptive}.
Each expert is typically responsible for modeling a distinct region of the input space.
A learnable routing network dynamically interpolates between these experts, producing weights that reflect their applicability to a given input.

\begin{figure}[t]
    \centering
    \includegraphics[width=0.9\linewidth]{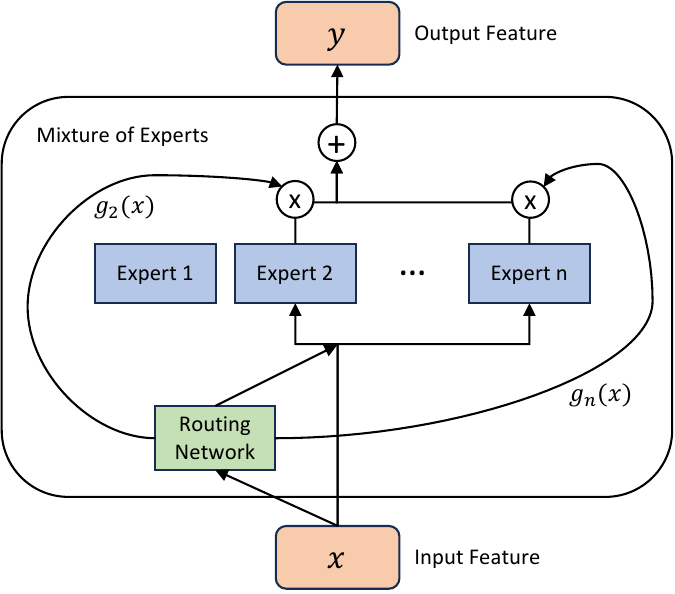}
    \caption{A conventional framework of classical mixture of experts.}
    \label{fig:classical_moe}
\end{figure}

Figure \ref{fig:classical_moe} illustrates a conventional framework of classical mixture of experts. Mathematically, given an input vector $\mathbf{x} \in \mathbb{R}^d$ and a set of $K$ experts $\{E_k\}_{k=1}^K$, a routing function $g(\mathbf{x})$ yields a probability distribution over the experts.
The output is computed as:
\begin{equation}
    \mathbf{y} = \sum_{k=1}^K g_k(\mathbf{x}) E_k(\mathbf{x})
\end{equation}
The routing network, typically implemented as a lightweight architecture, is trained jointly with the experts, enabling an end-to-end learning procedure.

\subsection{Parameterized Quantum Circuits}

The parameterized quantum circuit (PQC) \cite{benedetti2019parameterized} is a special kind of quantum circuit with parameters that can be optimized or learned iteratively.
The PQC comprises three parts, including data encoding, a parameterized layer, and quantum measurements.

Given a classical data $\mathbf{z} \in \mathbb{R}^{D}$ where $D$ is the data dimension, the data encoding circuit $U(\mathbf{z})$ is used to transform $\mathbf{z}$ into a quantum state $|\psi\rangle$.
The quantum state $|\psi\rangle$ is transformed via the parameterized layer $V(\theta)$ to a new state $|\psi\rangle$.
The parameterized layer is a sequence of quantum circuit operators with learnable parameters denoted as:
$V(\theta) = V_L(\theta_L) V_{L-1}(\theta_{L-1}) \dots V_{1}(\theta_{1})$,
where $L$ is the number of operators.
The learnable parameters can be updated via gradient-based \cite{mitarai2018quantum}, or gradient-free \cite{chen2022variational} algorithms.
The quantum measurements $H$ are used to retrieve the values of the quantum state for further processing.
Overall, the PQC can be formulated as:
\begin{equation}
    \langle H \rangle = \langle 0 | U^\dagger(\mathbf{z}) V^\dagger(\theta) H V(\theta) U(\mathbf{z}) | 0 \rangle
\end{equation}
where $H$ is a predefined observable.

PQC employs a hybrid quantum-classical approach to iteratively optimize the trainable parameters.
All learning methods take the training data as input and evaluate the performance of the model by comparing the predicted and ground-truth labels. 
Based on this evaluation, the methods update the model parameters for the next iteration and repeat the process until the model converges and achieves the desired performance. 

\section{Related Work}

The application of quantum computing methodologies to machine learning has been a rapidly evolving field of research in recent years. 
Early works focused on accelerating classical linear machine learning methods by utilizing the speedup from quantum algorithms. 
This foundational work demonstrated how quantum algorithms could be applied to traditional machine learning tasks, e.g., clustering \cite{lloyd2013quantum}, principal component analysis \cite{lloyd2014quantum}, least-squares fitting \cite{schuld2016prediction,kerenidis2020quantum}, and binary classification \cite{rebentrost2014quantum}. 
Prior works focused on using variational quantum algorithms or parameterized quantum circuits \cite{panella2011neural,mitarai2018quantum} to develop quantum neural networks. 
From this framework, Cong et al. \cite{cong2019quantum} introduced the quantum convolutional neural network, extending the properties of classical convolutional neural networks to quantum computing with fewer trainable parameters. 
Similarly, Bausch \cite{bausch2020recurrent} developed a quantum recurrent neural network by utilizing the structure of variational quantum eigensolver circuits. 
Furthermore, Huang et al. \cite{huang2021experimental} presented hybrid quantum generative adversarial networks for efficient data generation. 
In addition, Romero et al. \cite{romero2017quantum} introduced quantum autoencoders for dimensionality reduction of quantum states.

Despite their potential, most QML models lack scalability mechanisms that are critical for deployment on larger systems.
Classical deep learning has tackled similar issues using modular and sparse structures, notably the MoE frameworks \cite{jacobs1991adaptive,shazeer2017outrageously,fedus2022switch}.
MoEs enable specialization through multiple expert networks, dynamically routing inputs to a subset of them, which significantly improves model capacity and efficiency.
Integrating MoE principles into QML remains largely unexplored. 
A study has attempted to introduce modular quantum circuits \cite{tognini2025solving}.
However, the approach typically relies on classical control between routing and expert circuits.
Our proposed QMoE framework addresses this gap by introducing a quantum-native modular architecture that offers the potential for scalable QML systems.

\section{The Proposed Quantum Mixture of Experts}

\begin{figure*}[t]
\centering
\includegraphics[width=0.9\linewidth]{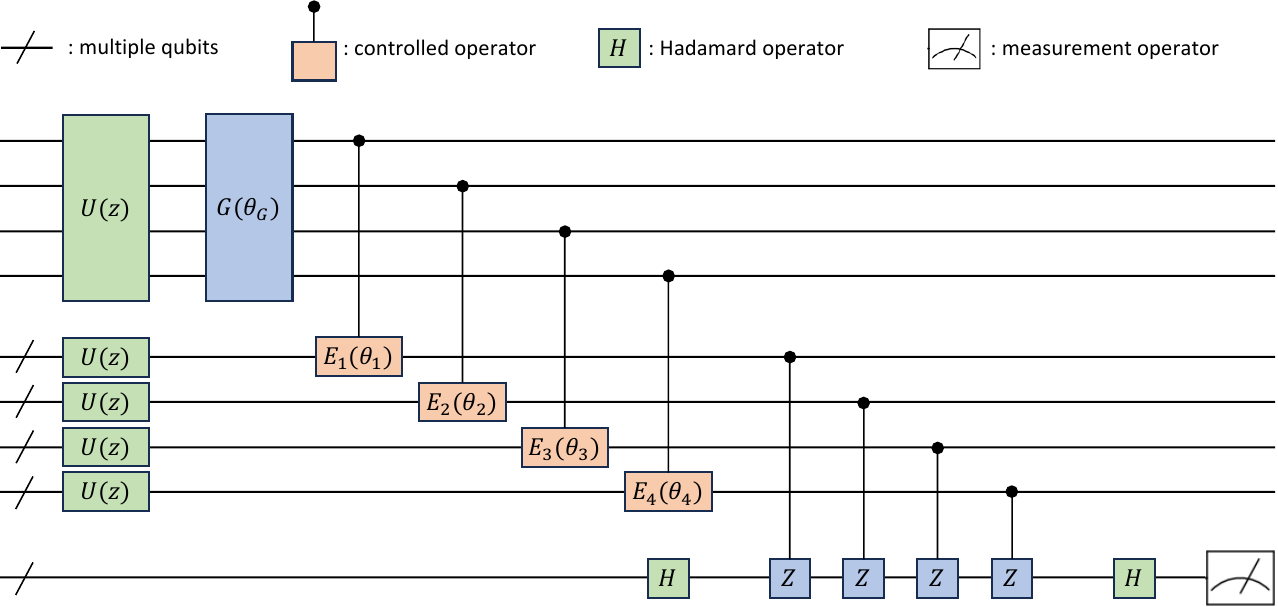}
\caption{
    The overall framework of the proposed quantum mixture of experts.
    The framework includes a classical data encoding circuit $U(\mathbf{z})$, a quantum routing circuit $G$, parameterized quantum experts $E_i$, and quantum states aggregation.
}
\label{fig:overall_framework}
\end{figure*}

Figure \ref{fig:overall_framework} illustrates the overall framework of the proposed quantum mixture of experts.
Inspired by the classical mixture of experts paradigm, our framework integrates quantum computing components to enable data-dependent routing, expert specialization, and output aggregation in a quantum setting.
Given a classical input $\mathbf{z} \in \mathbb{R}^D$, the data is encoded into a quantum state $|\psi\rangle$ via a data encoding circuit $U$.
Then, a learnable quantum routing circuit $G$ is introduced to select the parameterized quantum experts $E_i$ for quantum state transformation.
Finally, a set of transformed quantum states is aggregated for a projective measurement, yielding a classical output.
This output can be used for specific tasks such as classification or regression, enabling an end-to-end training via gradient-based methods.

\subsection{Quantum Routing Circuit}

\begin{figure}
    \centering
    \includegraphics[width=\linewidth]{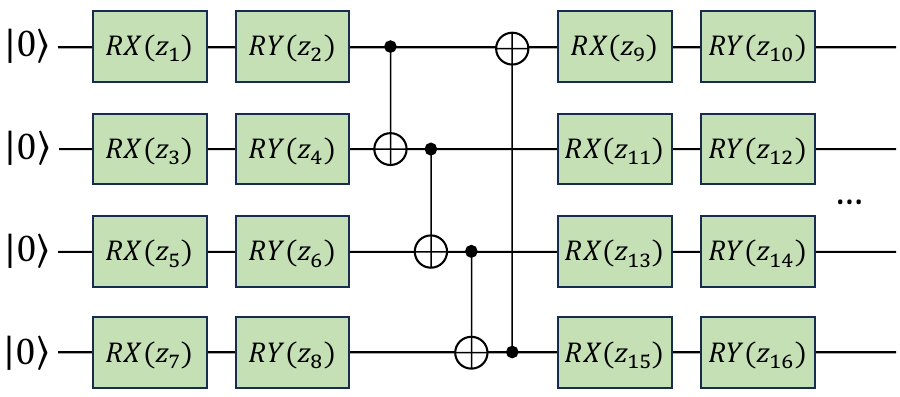}
    \caption{The classical data encoding circuits $U(\mathbf{z})$ include layers of rotation and controlled circuits with the input data elements as rotation parameters.}
    \label{fig:data_encoding}
\end{figure}

In classical neural networks, the routing network is a fundamental module to determine which experts should process a given input.
Analogously, in the proposed QMoE framework, the quantum routing circuit plays an essential role by dynamically selecting expert circuits based on learned routing decisions.
The input classical data $\mathbf{z}$ is first encoded into a quantum state via the encoding circuit $U(\mathbf{z})$, as shown in Figure \ref{fig:data_encoding}, which is applied to all routing and expert qubits in parallel.
Following this encoding, a parameterized quantum routing circuit $G(\theta_G)$ is applied to the routing.
This circuit learns a superposition over expert selection patterns through its trainable parameters $\theta_G$.
The output of the routing circuit determines which subset of experts $E_i(\theta_i)$ is activated via controlled operations.

\subsection{Parameterized Quantum Experts}

\begin{figure}[t]
    \centering
    \includegraphics[width=0.9\linewidth]{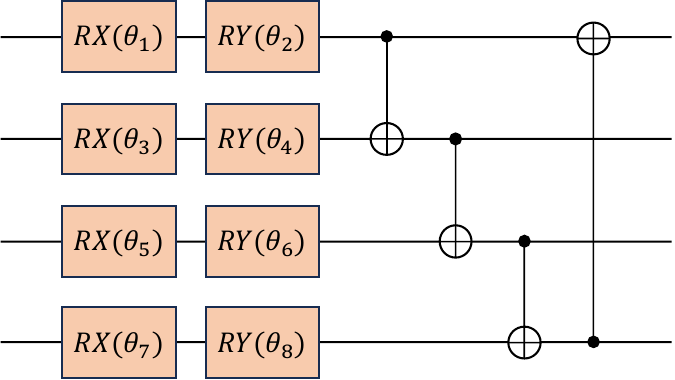}
    \caption{A conventional architecture of parameterized quantum circuits.}
    \label{fig:pqc}
\end{figure}

The classical MoEs rely on multiple specialized subnetworks or experts to learn different subtasks or data patterns.
Similarly, the proposed QMoE framework includes a set of parameterized quantum experts $\{E_i(\theta_i)\}_{i=1}^L$ where $L$ is the number of experts to model different patterns of the input quantum state.
These parameterized quantum experts are operated based on the controlled operators from the quantum routing circuit.
Each expert can be built from any parameterized quantum circuit architecture as shown in Figure \ref{fig:pqc}.
By training these experts jointly with the routing circuit, the QMoE learns both how to specialize experts and when to invoke them, similar to the modular learning behavior of classical MoEs.
This co-adaptation of routing and experts is fundamental to the scalable behavior of the proposed QMoE.
Moreover, it introduces sparsity and interpretability, as each expert tends to specialize in certain data characteristics or subtasks.

\begin{figure*}[t]
    \centering
    \includegraphics[width=0.7\linewidth]{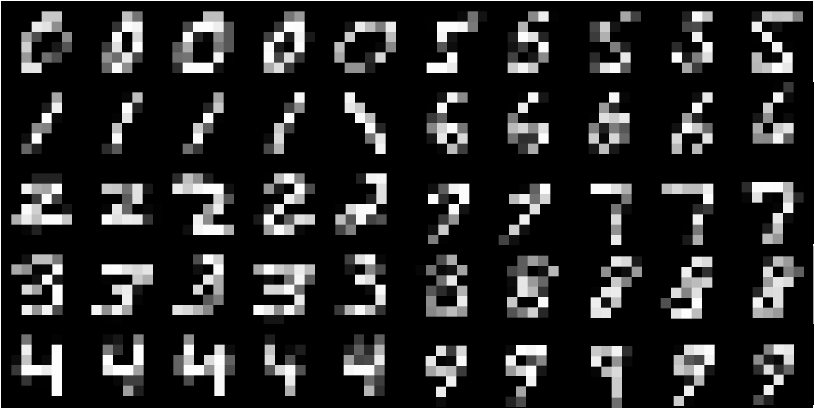}
    \caption{Data samples of the MNIST dataset.}
    \label{fig:mnist_samples}
\end{figure*}

\begin{table*}[t]
\centering
\caption{
Experimental accuracies (\%) on 2- and 4-class benchmarks.
}
\begin{tabular}{c|c|c|cccc}
\Xhline{2\arrayrulewidth}
\textbf{Compute Type} & \textbf{Method} & \textbf{Operators} & \textbf{MNIST-4} & \textbf{MNIST-2} & \textbf{Fashion-4} & \textbf{Fashion-2} \\
\hline
Quantum & Baseline & RX & 49.87 & 92.99 & 49.53 & 74.60 \\
Quantum & Baseline & RY & 49.27 & 93.04 & 49.13 & 75.35 \\
Quantum & Baseline & RX + RY & 51.05 & 94.89 & 52.58 & 78.35 \\
Quantum & Baseline & RX + RY + RZ & 51.47 & 94.68 & 53.65 & 79.94 \\
\hline
Quantum & \textbf{QMoE} & RX & 60.99 & 96.53 & 61.98 & 82.83 \\
Quantum & \textbf{QMoE} & RY & 60.29 & 97.13 & 60.44 & 82.47 \\
Quantum & \textbf{QMoE} & RX + RY & 62.92 & 98.62 & 63.47 & 83.94 \\
Quantum & \textbf{QMoE} & RX + RY + RZ & 61.26 & 97.29 & 63.81 & 84.46 \\
\Xhline{2\arrayrulewidth}
\end{tabular}
\label{tab:exp_results}
\end{table*}

\subsection{Quantum States Aggregation}

In classical settings, the outputs of expert subnetworks are typically combined via a weighted sum or voting scheme.
In the quantum setting, we introduce a quantum state aggregation process to complete the computation.
As the parameterized quantum experts are conditionally operated via controlled gates from the quantum routing circuit, a set of simple controlled operators is applied to aggregate the quantum states transformed by the quantum experts.
The final aggregated quantum state is measured to obtain the classical output of the quantum state, which is specific to the task.
In the training phase, the output is compared with the ground truth to update the learnable parameters via gradient-based methods, such as the parameter-shift method.

\section{Experimental Results}

\subsection{Experiment Setup}

\noindent
\textbf{Datasets}
Our proposed method is evaluated on four classification tasks. 
We use two benchmarks, the MNIST dataset \cite{lecun2010mnist} and the Fashion-MNIST dataset \cite{xiao2017fashion}. 
For MNIST, 4-class (0, 1, 2, 3) and 2-class (3, 6) tasks are created. 
For Fashion-MNIST, we use 4-class (t-shirt/top, trousers, pullover, dress) and 2-class (dress, shirt).
Each class includes 6,000 samples for training and 1,000 samples for testing.
The images are resized to a $8 \times 8$ resolution, and encoded via phase encoding with multiple rotation circuits. 
Samples of the MNIST benchmark inputs are illustrated in Figure \ref{fig:mnist_samples}.
We use the accuracy as an evaluation metric for experimental results.

\begin{table}[t]
\centering
\caption{
Ablation studies on different parameterized operators and different numbers of experts on the MNIST-4 benchmark.
}
\begin{tabular}{ccc}
\Xhline{2\arrayrulewidth}
\textbf{Operators} & \textbf{\# Experts} & \textbf{Accuracy (\%)} \\
\hline
 & 2 & 54.33 \\
RX & 3 & 58.54 \\
 & 4 & 60.99 \\
\hline
 & 2 & 54.15 \\
RY & 3 & 58.88 \\
 & 4 & 60.29 \\
\hline
 & 2 & 54.38 \\
RX + RY & 3 & 59.67 \\
 & 4 & 62.92 \\
\hline
 & 2 & 55.97 \\
RX + RY + RZ & 3 & 59.81 \\
 & 4 & 61.26 \\
\Xhline{2\arrayrulewidth}
\end{tabular}
\label{tab:ablation_studies}
\end{table}

\noindent
\textbf{Implementation}
The experiments for this work are conducted using quantum simulation for the quantum neural networks. 
The parameterized quantum circuits apply different learnable circuits, including RX, U2 (RX + RY), and U3 (RX + RY + RZ), and their efficiency is detailed in our ablation studies \ref{tab:ablation_studies}. 
Then, a non-learnable circuit, i.e., a non-learnable controlled-NOT (CNOT), is used. 
For the final comparison, we use four different experts in the proposed QMoE.
The training and testing of the framework are implemented using the TorchQuantum library \cite{wang2022torchquantum} for simulation on a Quadro RTX 8000 GPU. 
In our training process, we apply the Adam optimizer \cite{kingma2014adam} with a learning rate of $2 \times 10^{-3}$.

\subsection{Evaluation Results}

Table \ref{tab:exp_results} presents the classification accuracies across different datasets and model configurations.
We compare our QMoE models with baseline quantum models employing standard PQCs using different parameterized operators.
As shown in Table \ref{tab:exp_results}, the proposed QMoE framework consistently achieves higher accuracy compared to standard quantum baselines across all benchmarks.
On the MNIST-2 task, QMoE with RX+RY gates achieves 98.62\% accuracy, compared to 94.89\% for the best-performing quantum baseline.
Similar improvements are observed for the Fashion-2 dataset, where QMoE achieves 84.46\% accuracy with RX+RY+RZ gates, significantly higher than the 79.94\% of the corresponding quantum baseline.
On the MNIST-4 benchmark, QMoE improves the accuracy from 51.47\% to 62.92\% compared to the baseline.
The proposed QMoE also increases the accuracy from 53.65\% to 63.81\% on the Fashion-4 benchmark.
Notably, performance gains become more significant with the use of richer gate sets and task complexity. 

\subsection{Ablation Studies}

We further investigate the impact of different quantum gate configurations and the number of experts on the performance of QMoE using the MNIST-4 benchmark.
As shown in Table \ref{tab:ablation_studies}, we conduct ablation studies on four different parameterized operators, i.e., RX, RY, RX+RY, and RX+RY+RZ, each evaluated with two, three, and four experts.
As the number of experts increases, the accuracy improves across different operator settings.

\section{Conclusions}

In this paper, we have introduced quantum mixture of experts, a novel framework that integrates the classical mixture of experts paradigm into the quantum machine learning setting.
By combining a learnable quantum routing circuit with multiple parameterized quantum experts, the proposed framework enables input-adaptive specialization, modularity, and sparsity, which are essential for enhancing scalability and efficiency under noisy intermediate-scale quantum constraints.
We have validated our quantum mixture of experts on several quantum classification tasks using benchmark datasets such as MNIST and Fashion-MNIST, demonstrating consistent performance improvements over standard quantum neural network baselines. 
Additionally, ablation studies highlighted the flexibility of the proposed framework across various circuit configurations and expert counts.
The proposed quantum mixture of experts framework opens promising directions for building scalable, interpretable, and hardware-efficient quantum learning systems.

{
    \small
    \bibliographystyle{IEEEtran}
    \bibliography{IEEE_main}

\begin{thebibliography}{10}
\providecommand{\url}[1]{#1}
\csname url@samestyle\endcsname
\providecommand{\newblock}{\relax}
\providecommand{\bibinfo}[2]{#2}
\providecommand{\BIBentrySTDinterwordspacing}{\spaceskip=0pt\relax}
\providecommand{\BIBentryALTinterwordstretchfactor}{4}
\providecommand{\BIBentryALTinterwordspacing}{\spaceskip=\fontdimen2\font plus
\BIBentryALTinterwordstretchfactor\fontdimen3\font minus \fontdimen4\font\relax}
\providecommand{\BIBforeignlanguage}[2]{{%
\expandafter\ifx\csname l@#1\endcsname\relax
\typeout{** WARNING: IEEEtran.bst: No hyphenation pattern has been}%
\typeout{** loaded for the language `#1'. Using the pattern for}%
\typeout{** the default language instead.}%
\else
\language=\csname l@#1\endcsname
\fi
#2}}
\providecommand{\BIBdecl}{\relax}
\BIBdecl

\bibitem{preskill2012quantum}
J.~Preskill, ``Quantum computing and the entanglement frontier,'' \emph{arXiv preprint arXiv:1203.5813}, 2012.

\bibitem{preskill2018Quantum}
------, ``Quantum computing in the nisq era and beyond,'' \emph{Quantum}, vol.~2, p.~79, 2018.

\bibitem{boixo2018characterizing}
S.~Boixo, S.~V. Isakov, V.~N. Smelyanskiy, R.~Babbush, N.~Ding, Z.~Jiang, M.~J. Bremner, J.~M. Martinis, and H.~Neven, ``Characterizing quantum supremacy in near-term devices,'' \emph{Nature Physics}, vol.~14, no.~6, pp. 595--600, 2018.

\bibitem{basheer2020quantum}
A.~Basheer, A.~Afham, and S.~K. Goyal, ``Quantum $ k $-nearest neighbors algorithm,'' \emph{arXiv preprint arXiv:2003.09187}, 2020.

\bibitem{rebentrost2014quantum}
P.~Rebentrost, M.~Mohseni, and S.~Lloyd, ``Quantum support vector machine for big data classification,'' \emph{Physical review letters}, vol. 113, no.~13, p. 130503, 2014.

\bibitem{horn2001method}
D.~Horn and A.~Gottlieb, ``The method of quantum clustering,'' \emph{Advances in neural information processing systems}, vol.~14, 2001.

\bibitem{horn2001algorithm}
------, ``Algorithm for data clustering in pattern recognition problems based on quantum mechanics,'' \emph{Physical review letters}, vol.~88, no.~1, p. 018702, 2001.

\bibitem{nguyen2023quantum}
X.~B. Nguyen, H.~Churchill, K.~Luu, and S.~U. Khan, ``Quantum vision clustering,'' \emph{arXiv preprint arXiv:2309.09907}, 2023.

\bibitem{nguyen2024qclusformer}
X.-B. Nguyen, H.-Q. Nguyen, S.~Y.-C. Chen, S.~U. Khan, H.~Churchill, and K.~Luu, ``Qclusformer: A quantum transformer-based framework for unsupervised visual clustering,'' in \emph{2024 IEEE International Conference on Quantum Computing and Engineering (QCE)}, vol.~2.\hskip 1em plus 0.5em minus 0.4em\relax IEEE, 2024, pp. 347--352.

\bibitem{ezhov2000quantum}
A.~A. Ezhov and D.~Ventura, ``Quantum neural networks,'' in \emph{Future Directions for Intelligent Systems and Information Sciences: The Future of Speech and Image Technologies, Brain Computers, WWW, and Bioinformatics}.\hskip 1em plus 0.5em minus 0.4em\relax Springer, 2000, pp. 213--235.

\bibitem{zhou2023quantum}
M.-G. Zhou, Z.-P. Liu, H.-L. Yin, C.-L. Li, T.-K. Xu, and Z.-B. Chen, ``Quantum neural network for quantum neural computing,'' \emph{Research}, vol.~6, p. 0134, 2023.

\bibitem{gupta2020quantum}
B.~Gupta and S.~Dhawan, ``Quantum neural network (qnn) research a scientometrics assessment of global publications during 1990-2019.'' \emph{International Journal of Information Dissemination \& Technology}, vol.~10, no.~3, 2020.

\bibitem{dendukuri2019defining}
A.~Dendukuri, B.~Keeling, A.~Fereidouni, J.~Burbridge, K.~Luu, and H.~Churchill, ``Defining quantum neural networks via quantum time evolution,'' \emph{arXiv preprint arXiv:1905.10912}, 2019.

\bibitem{dendukuri2018image}
A.~Dendukuri and K.~Luu, ``Image processing in quantum computers,'' \emph{arXiv preprint arXiv:1812.11042}, 2018.

\bibitem{nguyen2024hierarchical}
X.-B. Nguyen, H.-Q. Nguyen, H.~Churchill, S.~U. Khan, and K.~Luu, ``Hierarchical quantum control gates for functional mri understanding,'' in \emph{2024 IEEE Workshop on Signal Processing Systems (SiPS)}.\hskip 1em plus 0.5em minus 0.4em\relax IEEE, 2024, pp. 159--164.

\bibitem{nguyen2025diffusion}
H.-Q. Nguyen, X.~B. Nguyen, S.~Y.-C. Chen, H.~Churchill, N.~Borys, S.~U. Khan, and K.~Luu, ``Diffusion-inspired quantum noise mitigation in parameterized quantum circuits,'' \emph{Quantum Machine Intelligence}, vol.~7, no.~1, p.~55, 2025.

\bibitem{nguyen2024quantum}
H.-Q. Nguyen, X.-B. Nguyen, H.~Churchill, A.~K. Choudhary, P.~Sinha, S.~U. Khan, and K.~Luu, ``Quantum-brain: Quantum-inspired neural network approach to vision-brain understanding,'' \emph{arXiv preprint arXiv:2411.13378}, 2024.

\bibitem{biamonte2017quantum}
J.~Biamonte, P.~Wittek, N.~Pancotti, P.~Rebentrost, N.~Wiebe, and S.~Lloyd, ``Quantum machine learning,'' \emph{Nature}, vol. 549, no. 7671, pp. 195--202, 2017.

\bibitem{du2020expressive}
Y.~Du, M.-H. Hsieh, T.~Liu, and D.~Tao, ``Expressive power of parametrized quantum circuits,'' \emph{Physical Review Research}, vol.~2, no.~3, p. 033125, 2020.

\bibitem{shazeer2017outrageously}
N.~Shazeer, A.~Mirhoseini, K.~Maziarz, A.~Davis, Q.~Le, G.~Hinton, and J.~Dean, ``Outrageously large neural networks: The sparsely-gated mixture-of-experts layer,'' \emph{arXiv preprint arXiv:1701.06538}, 2017.

\bibitem{fedus2022switch}
W.~Fedus, B.~Zoph, and N.~Shazeer, ``Switch transformers: Scaling to trillion parameter models with simple and efficient sparsity,'' \emph{Journal of Machine Learning Research}, vol.~23, no. 120, pp. 1--39, 2022.

\bibitem{jacobs1991adaptive}
R.~A. Jacobs, M.~I. Jordan, S.~J. Nowlan, and G.~E. Hinton, ``Adaptive mixtures of local experts,'' \emph{Neural computation}, vol.~3, no.~1, pp. 79--87, 1991.

\bibitem{benedetti2019parameterized}
M.~Benedetti, E.~Lloyd, S.~Sack, and M.~Fiorentini, ``Parameterized quantum circuits as machine learning models,'' \emph{Quantum Science and Technology}, vol.~4, no.~4, p. 043001, 2019.

\bibitem{mitarai2018quantum}
K.~Mitarai, M.~Negoro, M.~Kitagawa, and K.~Fujii, ``Quantum circuit learning,'' \emph{Physical Review A}, vol.~98, no.~3, p. 032309, 2018.

\bibitem{chen2022variational}
S.~Y.-C. Chen, C.-M. Huang, C.-W. Hsing, H.-S. Goan, and Y.-J. Kao, ``Variational quantum reinforcement learning via evolutionary optimization,'' \emph{Machine Learning: Science and Technology}, vol.~3, no.~1, p. 015025, 2022.

\bibitem{lloyd2013quantum}
S.~Lloyd, M.~Mohseni, and P.~Rebentrost, ``Quantum algorithms for supervised and unsupervised machine learning,'' \emph{arXiv preprint arXiv:1307.0411}, 2013.

\bibitem{lloyd2014quantum}
------, ``Quantum principal component analysis,'' \emph{Nature Physics}, vol.~10, no.~9, pp. 631--633, 2014.

\bibitem{schuld2016prediction}
M.~Schuld, I.~Sinayskiy, and F.~Petruccione, ``Prediction by linear regression on a quantum computer,'' \emph{Physical Review A}, vol.~94, no.~2, p. 022342, 2016.

\bibitem{kerenidis2020quantum}
I.~Kerenidis and A.~Prakash, ``Quantum gradient descent for linear systems and least squares,'' \emph{Physical Review A}, vol. 101, no.~2, p. 022316, 2020.

\bibitem{panella2011neural}
M.~Panella and G.~Martinelli, ``Neural networks with quantum architecture and quantum learning,'' \emph{International Journal of Circuit Theory and Applications}, vol.~39, no.~1, pp. 61--77, 2011.

\bibitem{cong2019quantum}
I.~Cong, S.~Choi, and M.~D. Lukin, ``Quantum convolutional neural networks,'' \emph{Nature Physics}, vol.~15, no.~12, pp. 1273--1278, 2019.

\bibitem{bausch2020recurrent}
J.~Bausch, ``Recurrent quantum neural networks,'' \emph{Advances in neural information processing systems}, vol.~33, pp. 1368--1379, 2020.

\bibitem{huang2021experimental}
H.-L. Huang, Y.~Du, M.~Gong, Y.~Zhao, Y.~Wu, C.~Wang, S.~Li, F.~Liang, J.~Lin, Y.~Xu \emph{et~al.}, ``Experimental quantum generative adversarial networks for image generation,'' \emph{Physical Review Applied}, vol.~16, no.~2, p. 024051, 2021.

\bibitem{romero2017quantum}
J.~Romero, J.~P. Olson, and A.~Aspuru-Guzik, ``Quantum autoencoders for efficient compression of quantum data,'' \emph{Quantum Science and Technology}, vol.~2, no.~4, p. 045001, 2017.

\bibitem{tognini2025solving}
P.~A.~X. Tognini, L.~Banchi, and G.~De~Palma, ``Solving mnist with a globally trained mixture of quantum experts,'' \emph{arXiv preprint arXiv:2505.14789}, 2025.

\bibitem{lecun2010mnist}
Y.~LeCun, C.~Cortes, and C.~Burges, ``Mnist handwritten digit database,'' \emph{ATT Labs [Online]. Available: http://yann.lecun.com/exdb/mnist}, vol.~2, 2010.

\bibitem{xiao2017fashion}
H.~Xiao, K.~Rasul, and R.~Vollgraf, ``Fashion-mnist: a novel image dataset for benchmarking machine learning algorithms,'' \emph{arXiv preprint arXiv:1708.07747}, 2017.

\bibitem{wang2022torchquantum}
H.~Wang, Z.~Liang, J.~Gu, Z.~Li, Y.~Ding, W.~Jiang, Y.~Shi, D.~Z. Pan, F.~T. Chong, and S.~Han, ``Torchquantum case study for robust quantum circuits,'' in \emph{Proceedings of the 41st IEEE/ACM International Conference on Computer-Aided Design}, 2022, pp. 1--9.

\bibitem{kingma2014adam}
D.~P. Kingma, ``Adam: A method for stochastic optimization,'' \emph{arXiv preprint arXiv:1412.6980}, 2014.

\end{thebibliography}
}

\end{document}